\documentclass[12pt,preprint]{aastex}

\shorttitle{Chandra Observations of NGC 4410A}
\shortauthors{Smith et al.}

\begin{document}


\title{Chandra Observations of the\\
Interacting NGC 4410 Galaxy Group}


\author{Beverly J. Smith}
\affil{Department of Physics and Astronomy,
East Tennessee State University, Johnson City TN  37614}
\email{smithbj@etsu.edu}

\author{Michael Nowak}
\affil{Massachusetts Institute of Technology}
\email{mnowak@alum.mit.edu}

\author{Megan Donahue}
\affil{Space Telescope Science Institute}
\email{donahue@stsci.edu}

\and

\author{John Stocke}
\affil{CASA, University of Colorado}
\email{stocke@casa.colorado.edu}



\begin{abstract}

We present high resolution X-ray imaging data from the ACIS-S instrument
on the Chandra
telescope of the nearby
interacting
galaxy group NGC 4410.  Four galaxies in the inner portion
of this group are clearly detected by Chandra, including 
the peculiar low luminosity radio galaxy NGC 4410A.
In addition to a nuclear point source,
NGC 4410A contains
diffuse X-ray emission, including an X-ray ridge extending out to about 
12$''$ (6 kpc)
to the northwest of the
nucleus.  This ridge is
coincident with an arc of optical emission-line gas, which 
has previously
been shown to have optical line ratios consistent with shock ionization.
This structure may be due to an expanding superbubble of hot gas 
caused by supernovae and stellar winds
or by the active nucleus.
The Chandra observations also 
show four or five possible compact
ultra-luminous
X-ray (ULX) sources (L$_x$ $\ge$ 10$^{39}$ erg~s$^{-1}$) associated
with NGC 4410A.
At least one of 
these candidate ULXs appears to have a radio counterpart, suggesting
that it may be due to an X-ray binary with a stellar-mass black hole,
rather than an intermediate mass black hole. 
In addition, a 
faint 
diffuse intragroup X-ray component has been
detected between the galaxies (L$_x$ $\sim$ 10$^{41}$ erg~s$^{-1}$).
This supports the hypothesis that
the NGC 4410 group is
in the process of evolving via
mergers from a spiral-dominated group (which typically
have no X-ray-emitting intragroup gas) 
to an elliptical-dominated group 
(which often have a substantial intragroup medium).

\end{abstract}




\section{Introduction}

There has been increasing evidence in recent years that 
most
elliptical galaxies are the
result of the merging of disk galaxies 
(e.g., \citet{b85, bh92}).  
The optical counterparts of luminous radio galaxies are generally
elliptical or elliptical-like galaxies
\citep{ow91}, which
suggests
that the creation of radio jets may somehow be linked to the
merger process 
\citep{wc95, ms96}.
Once a radio jet is formed, its subsequent evolution may
be affected by further galaxy interactions and mergers.
For example, it may be distorted 
by
motion through an intracluster medium 
\citep{m72},
by an encounter with interstellar matter in either a
host galaxy 
\citep{dy81}
or a companion galaxy 
\citep{s85, bc93}
or by a galaxy in the process of merging with the host
galaxy 
\citep{vb86}.
In addition, impacts between jets
and interstellar matter may trigger star formation in some cases 
\citep{dy81, vb85}.

It is difficult to study these processes because
most elliptical galaxies, and most radio galaxies, reside 
in
rich clusters where the merger events took place very long ago.
Some ellipticals, however, are found in poor groups
in the current epoch 
(see, e.g., \citet{mz98}), which are
only now undergoing the merger events that took place long ago in rich
clusters. Poor, compact groups of galaxies, therefore, are the locations
best suited for investigating the merger process.
\citet{mz98}
have found hot, X-ray halos around dominant ellipticals in
poor galaxy groups, however,
usually the elliptical
already looks completely formed and the gaseous halo is relaxed. Evidently,
the actual merger events themselves happen so quickly that there are very
few nearby examples. If a few examples were found, however, they would
offer a complex set of observational constraints since the merging
participants could include gas and stars from a remnant disk galaxy,
new star
formation driven by the merger, inception of AGN activity in one or more
nuclei, and jet driven formation by the newly created radio jet. We have
discovered a nearby example of a likely merger in progress: NGC 4410.

The NGC 4410 group contains
about a dozen members, with five major galaxies in its inner portions,
at least four of which appear to be strongly interacting
(see Figure 1).
This group is quite nearby
(97 Mpc, for
H$_0$ = 75 km s$^{-1}$ Mpc$^{-1}$), and is well-studied
at radio and optical wavelengths 
\citep{h86, s00, d02}.
The four brightest galaxies in the inner group and the four brightest
in the outer group are classified as spiral or S0 galaxies 
in the NASA/IPAC Extragalactic
Database (NED\footnote{The NASA/IPAC Extragalactic
Database at http://nedwww.ipac.caltech.edu
is operated by the Jet Propulsion Laboratory,
California Institute of Technology,
under contract with the National
Aeronautics and Space Administration}).
The most luminous member at optical wavelengths, NGC 4410A, 
classified as an Sab? peculiar
galaxy by \citet{deV93},
is a low
luminosity radio galaxy with a 
very distorted double-lobed radio morphology 
\citep{h86, s00}.
The pecular radio structure
strongly suggests that the radio jet has been affected
by motion relative to interstellar or intracluster gas
\citep{sb87, s00}.
NGC 4410A
has a peculiar ring-like optical morphology, with six extremely
luminous H~II regions (L$_{H\alpha}$ = $10^{40}$--$10^{41}~{\rm
erg~sec^{-1}}$) lying along this structure (Figure 2; \citep{d02}).
At present, it is not clear whether the star formation in NGC~4410A
was triggered by pressure from the radio lobes on the ambient medium, or by
a gravitational interaction or collision with NGC~4410B.
The western portion of this ring has optical line ratios consistent
with shock ionization \citep{d02}.
NGC 4410A has been
detected in both 21 cm H~I and 2.6mm CO (M$_{\rm HI}$ = 1.3~$\times$
10$^9$ M$_{\sun}$ and M$_{\rm H_2}$ = 3.9 $\times$ 10$^9$ M$_{\sun}$, assuming
the standard Galactic conversion factor) \citep{s00}.  
A prominent H~I tail has been
discovered, aligned along a stellar tail extending 1.7 arcmin (50 kpc) to
the southeast of the NGC 4410A/NGC 4410B pair \citep{s00}.
Both the radio galaxy NGC 4410A and its nearby companion
NGC 4410B are classified as Low Ionization Nuclear Emission Region (LINER) 
galaxies \citep{mb93, d02},
and NGC 4410D and F also show optical emission lines \citep{sb87}.

Clues to the history of the NGC 4410 system may come from X-ray data.  The
NGC 4410A group has previously 
been mapped in X-rays with both the High Resolution
Imager (HRI) and the Position Sensitive Proportional Counter (PSPC)
instruments on ROSAT 
\citep{t99, s00}.
The global PSPC
spectrum has been deconvolved into two components: a power-law spectrum,
contributing 2/3 of the X-ray radiation, and a 1\,keV thermal component,
contributing the remaining light 
\citep{t99}.
The HRI map
shows a point source associated with the nucleus of NGC 4410A and a diffuse
halo extending 10$''$ to the southeast, towards the most luminous H~II
region.  The halo contributes approximately 1/3 of the total light in the
HRI map, supporting its association with the thermal component in the PSPC
spectral decomposition.  The lower resolution PSPC map shows a faint
tail-like feature extending eastwards from the NGC 4410A+B pair towards NGC
4410C, anti-coincident with the southeastern H~I tail.  From the ROSAT
data, it was
not clear whether this X-ray emission is due to shocked gas in
the tidal bridge, 
as has been suggested for X-ray-emitting gas associated
with a tidal tail in the radio galaxy Fornax A \citep{mf98},
intragroup gas that is coincidently superposed on the stellar
bridge, or simply background or foreground sources.  

\section{Observations}

To address these issues, we have made higher resolution
X-ray observations of the NGC 4410 galaxy 
group in imaging mode with the backside-illuminated 
Advanced CCD Imaging Spectrometer (ACIS) 
S3 chip on the Chandra 
X-ray telescope \citep{w02}.
The nucleus of NGC 4410A was placed near the nominal aimpoint
of this chip, and the roll angle during the observations
was 203$^{\circ}$ west of north.
The ACIS chips have 1024 $\times$ 1024 0\farcs492 pixels.
To
reduce pile-up, we only read out 640 chip rows, giving 
an observed field of view of 5\farcm2 $\times$ 8\farcm4.
The other four galaxies in the
inner group (NGC 4410B, C, D, and F) also fell on this 
chip.
The CCD temperature was $-$120$^{\circ}$ C, the frame time
was 2.1 seconds, and the events were telemetered in Faint mode.
In addition to the S3 chip (the main focus of this paper),
the S1, S2, S4, I3, and I4 chips of ACIS were also read out.

Initial data reductions were done using the CIAO v2.2.1 
software\footnote{CIAO is the Chandra Interactive Analysis
of Observations data analysis system package 
(http://cxc.harvard.edu/ciao)}.
After removing time intervals with high X-ray background,
the useable exposure time was 34.3 ksec.
Only events with ASCA grades 0, 2, 3, 4, and 6 were included,
and standard bad pixels and columns were removed.
The energy range was restricted to 0.3 $-$ 8 keV,
and then further divided into three ranges:
0.3 $-$ 1 keV, 1 $-$ 2.5 keV, and 2.5 $-$ 8 keV.
The binned images were adaptively smoothed 
using the CIAO routine
csmooth, using a minimal significance of S/N = 2.5 and a maximal
significance of S/N = 5.
The smoothed images were then divided by a similarly-smoothed
exposure map to convert into
physical units and to remove instrumental artifacts.
We experimented with exposure maps created using different parameters,
including both monochromatic maps at various energies and weighted maps.
In all cases, the basic morphology of the extended emission in
the group remained the same, although the absolute flux level varied.
For our final 0.3 $-$ 8 keV map, we used an exposure map weighted
by the spectrum of a 57\farcs5 $\times$ 40$''$ region centered on
the NGC 4410A+B complex.  For the final 0.3 $-$ 1 keV,
1 $-$ 2.5 keV, and 2.5 $-$ 8 keV maps, we used
monochromatic 0.8 keV, 1.5 keV, and 3 keV exposure maps, respectively.

\section{Results}

\subsection{Morphology}

In Figure 3, we provide 
the final 0.3 $-$ 8 keV ACIS-S S3 Chandra map (greyscale), with 
optical contours from the 
Digitized Sky Survey\footnote{The Digitized
Sky Survey was produced at the
Space Telescope Science
Institute under U.S. Government
grant NAG W-2166.
These images are
digitized versions of
photographic plates from the
Second Palomar Observatory
Sky Survey (POSS-II) made by
the California Institute
of Technology with funds from
the National Science Foundation,
the National Geographic Society, the Sloan
Foundation, the Samuel Oshin Foundation, and the Eastman
Kodak Corporation}
(DSS) superposed.
This figure shows that
four galaxies in the inner part of the NGC 4410 group 
were detected by Chandra: NGC 4410A+B, NGC 4410C, and NGC 4410D.
Several point sources are also visible in this map,
including two sources in the NGC 4410A+B$-$C bridge.
The more northeastern of these sources (at 
12$^{\rm h}$~26$^{\rm m}$~33.5$^{\rm s}$, 
9$^{\circ}$~1$'$~45$''$ (J2000)),
is particularly bright, with a total
of 175 0.3 $-$ 8 keV counts, and is slightly extended
($\sim$2$''$ $\times$ 3$''$).

This map also shows faint extended emission in the group, concentrated
around NGC 4410A+B, and extending
out past NGC 4410C.
Over the mapped portion of the
group, for the diffuse X-ray emission outside of the
main optical galaxies,
the average Chandra 0.3 - 8 keV counts are 9 $\pm$ 2 
$\times$ 10$^{-10}$ 
photons s$^{-1}$ cm$^{-2}$ pixel$^{-1}$, with the uncertainty
dominated by the uncertainty in the exposure map (the statistical noise
is only $\sim$0.3 $\times$ 10$^{-10}$
photons s$^{-1}$ cm$^{-2}$ pixel$^{-1}$).
Assuming a typical energy per photon of 0.8 keV,
this corresponds to 5 $\pm$ 1 $\times$ 10$^{-18}$
erg s$^{-1}$ cm$^{-2}$ arcsec$^{-2}$,
consistent with the upper limit from the ROSAT data
\citep{s00}.

Another concentration of extended emission
is visible to the northeast of NGC 4410D, at $\sim$12$^{\rm h}$~
26$^{\rm m}$~47$^{\rm s}$, 9$^{\circ}$,~4$'$~15$''$.
This region of X-ray emission is also present in the ROSAT maps of 
\citet{t99}.  In the higher resolution Chandra map, it is resolved
into four point sources, surrounded by extended emission.
One of these X-ray sources has a possible optical counterpart 
in the DSS image.
These point sources may be background QSOs or radio galaxies
in a distant cluster rich in hot X-ray-emitting intracluster gas.

In Figure 4, the low energy (0.3 $-$ 1 keV) map is presented,
again superposed on
the DSS image.
The extended emission in this map has a morphology similar to
that in the total 0.3 $-$ 8 keV map.
Extended emission is visible around
NGC 4410A+B extending up to NGC 4410C. The second
concentration
to the northeast of NGC 4410D is also visible.
In this low energy map, a faint source is visible near NGC 4410F.

In the middle energy (1 $-$ 2.5 keV) map (Figure 5), the morphology
of the central 
region is similar to that seen in the
low energy map.  In the northeast, however, the morphology
is somewhat different.  In particular, the point source
northeast of NGC 4410D
with the possible optical counterpart is no longer
visible.

At even higher energies (2.5 $-$ 8 keV; Figure 6),
the nucleus of NGC 4410A is quite
bright, as is the more northeastern point
source in the NGC 4410A+B-C bridge.
NGC 4410C and D, however, are quite faint, and NGC 4410F is 
undetected.
This map shows a different extended morphology
in the central regions 
than the lower energy maps, although we note that the highest
energies are the most likely to be contaminated by background.
Faint diffuse emission is visible between NGC 4410A+B and C,
and north of NGC 4410D.   If this is real, it implies that the 
intracluster gas in this region is hotter than in the rest of the group.

The Chandra position of the nucleus of NGC 4410A
agrees with the radio continuum
peak position from \citet{h86} within 0.1$''$.
The 4.9 GHz radio continuum map of 
the NGC 4410 group from \citet{h86} is superposed on
the 0.3 $-$ 8 keV Chandra map in Figure 7.   Note the apparent
anti-coincidence of the extended X-ray and radio emission 
in the southeast.  The radio continuum lobe lies to
the south, while the X-ray emission extends approximately
east-west.  
The Chandra field of view does not
extend beyond the radio lobe, so it is not known whether
the radio lobe is surrounded by X-ray-emitting gas.
The northern radio lobe, seen in the lower resolution
map of \citet{s00}, is too faint to be visible in this higher
resolution \citet{h86}  map.

An expanded view of the portion of the
0.3 $-$ 8 keV map around NGC 4410A+B is shown in
Figure 8, and, in Figure 9, a red archival Hubble Space Telescope
image is superposed as contours on the unsmoothed Chandra map.
In Figure 10, the \citet{d02} H$\alpha$+$[$N~II$]$ map is superposed
on the Chandra map.
The bright X-ray peak is associated with the 
NGC 4410A nucleus, while the secondary source to the east is
coincident with the NGC 4410B nucleus.  A ridge
of X-ray emission extending $\sim$12$''$ (6 kpc) to
the northwest of NGC 4410A is also visible, coincident with the ionized
gas in the northwestern portion of the optical loop (Figure 10).
As noted earlier, this gas has optical line ratios indicative
of shock ionization \citep{d02}. 
Note that the 3$''$ extension to the southwest of the NGC 4410A nucleus
seen in the optical line map is also present in X-rays.
Note also that the optically-bright knots to the southeast and northeast
of the nucleus of NGC 4410A are not visible as discrete sources
in the X-ray map.  These sources are known to be luminous
H~II regions \citep{d02}.

\subsection{X-Ray Spectra}

Spectra of
NGC 4410A $-$ D, 
the brightest bridge source in the NGC 4410A+B$-$C
bridge, and the diffuse intragroup gas 
were extracted from the data using CIAO,
and spectral analysis was accomplished using
the ISIS data reduction package \citep{hd00}.
Results of spectral fitting are given in Table 1 and Figures 11 and 12.
Unless otherwise noted,
the 0.5 $-$ 5 keV range was used for fitting the spectra, eliminating
the low S/N high energy portion plus the lower energies
where the calibration is somewhat uncertain.
The quoted uncertainties are 90$\%$ confidence level.
Before fitting the spectra, 
the data were rebinned into 20 counts bin$^{-1}$ and
the recent CIAO 
{\it apply$\_$acisabs} 
correction
for contamination of the ACIS chips was applied.
To estimate background counts, the deep ACIS observations
of blank fields provided by M. Markevitch were used
(file acis7sD2000-12-01bkgrndN0002.fits).  

In no case were we able 
to successfully
constrain the absorption column, the normalization, and the power law
index or temperature simultaneously.
For NGC 4410A, we divided the source into three radial regimes: 
$<$ 1$''$, 1$''$ $-$ 5$''$, and 5$''$ $-$ 10$''$.
The nucleus of NGC 4410A is relatively bright (1256 counts
in a 4.25 pixel radius), so we also included
the effects of pile-up for the central region.
When fixing the absorption to a nominal
value of 5 $\times$ 10$^{20}$ cm$^{-2}$ 
(approximately equal to the sum
of the 1.7 $\times$ 10$^{20}$ cm$^{-2}$ Galactic absorption
from
\citet{dl90}
and the intrinsic HI column densities
from \citet{s00}),
for the inner 1$''$ 
we obtained
an adequate fit to a power law with photon index $\Gamma$ $\sim$ 2 (see 
Table 1).
Conversely, by fixing the photon index to this value, we obtained 
column densities consistent with the nominal value.
Increasing the photon index increases the fitted column density, still
providing good fits with reasonable column density
up to $\Gamma$ $\sim$ 2.3. 
Models without pile-up give similar results.

For the 1$''$ $-$ $5''$ and 5$''$ $-$ 10$''$ annuli around NGC 4410A,
the spectra can be fit with Mekal functions, with temperatures
kT = 0.62 $\pm$ $^{0.08}_{0.07}$ keV
and 0.54 $\pm$ $^{0.12}_{0.14}$ keV and abundances 
of 0.06 $\pm$ $^{0.06}_{0.03}$ and
0.11 $\pm$ 
$^{1.16}_{0.09}$ solar, respectively.
For fitting the inner and outer annuli, we only included photons with energies 0.6 $-$ 1.8 keV
and 0.6 $-$ 1.2 keV; at higher energies, the counts are consistent with the background.

The nuclei of B, C, and D had very few counts, so it was not possible
to do detailed fitting of their spectral shapes.
For these objects,
we fixed both the absorbing column and the spectral parameters, and checked
for consistency.  All three spectra are consistent with bremsstrahlung
spectra, and are consistent with both
the Galactic absorption alone or with the nominal internal absorption included.
The NGC 4410B spectrum appears relatively soft, consistent with a kT = 0.2 keV
temperature, while NGC 4410C and D are consistent with kT = 0.5 $-$ 1 keV.

For the bridge source (named `b1' in Table 1), we found a good fit to the photon
index $\Gamma$ = 2.2 $\pm$ $^{0.3}_{0.2}$ 
with the nominal extinction.
If it is at the distance of NGC 4410, it has a luminosity
of 4 $\times$ 10$^{40}$ erg~s$^{-1}$.  At that distance,
the angular extent of the source corresponds to a size 
of $\sim$1~kpc $\times$ 1.5~kpc.

To obtain the spectrum and luminosity of the diffuse intragroup medium,
we extracted counts in a 6\farcm4 $\times$ 4\farcm6 (180 $\times$ 130 kpc)
diameter ellipse
aligned east-west,
centered at 12$^{\rm h}$ 26$^{\rm m}$ 32.7$^{\rm s}$, 9$^{\circ}$ 1$'$ 45\farcs5 (J2000),
excluding 51$''$, 21$''$, and 21$''$ diameter circular regions 
around NGC 4410A+B, NGC 4410C, and NGC 4410D,
respectively, as well as 5$''$ radius regions around 
the bright bridge source and 11 point sources visible in
the field (including the point sources discussed in Section 3.3).
This ellipse does not cover the concentration of diffuse 
gas northeast of NGC 4410D,
which may be background emission.
The extracted spectrum is displayed in Figure 12a (solid line), along with
the background counts from the 
Markevitch deep fields, reprojected and scaled to 
our observations (dotted line).  
This comparison shows that 
essentially all of the observed counts
below 2 keV are due to the background, 
as are the two apparent `spectral features'
seen in the raw data at 1.8 and 2 keV 
(see also Figure 4 in \citet{strick02}).
This background
spectrum is consistent with, but higher signal/noise than,
spectra for selected regions
at the edge of our field where little diffuse emission
is seen.

The background-subtracted spectrum for the diffuse
intragroup gas is shown in Figure 12b.  
Excluding energies above 1.8 keV and below 0.6 keV, 
where the background subtraction and calibration are more uncertain,
fitting this spectrum with 
a Mekal function gives kT = 0.69 $\pm$ $^{0.15}_{0.16}$ 
keV and a metallicity
of 0.025 $\pm$ $^{0.023}_{0.024}$ solar.
The total 0.3 $-$ 8 keV luminosity of this 
diffuse gas is 10$^{41}$ erg~s$^{-1}$.

\subsection{Point Sources}

Depending upon the exact spectral shape, our observations
are sensitive to point sources 
at the distance of NGC 4410
with 0.5 $-$ 8 keV luminosities
of $\ge$ 10$^{39}$ erg~s$^{-1}$. 
To search for such sources, we
used the CIAO {\it wavedetect} routine.  
In addition to the galactic 
nuclei and the bright bridge source, we detected 8 candidate sources
in the vicinity of NGC 4410
(see Figure 13, where they are marked on the optical
DSS image).
These sources are listed in Table 2, along with counts
in six different energy bands.  
None of these sources show evidence of variability, although
the brightest only has 31 counts.

The first three energy bands listed in Table 2,
E$_1$ (0.5 $-$ 0.95 keV), E$_2$ (0.95 $-$ 1.5 keV), and E$_3$ (1.5 $-$ 8 keV),
are expected to have equal counts for a $\Gamma$ = 2 power law
and nominal absorption (N$_H$ $\sim$ 5 $\times$ 10$^{20}$ cm$^{-2}$).
This table shows that source
p8 has a hard spectrum, with no photons below 4 keV.
Source p3 also has an over-abundance of hard photons relative to
a $\Gamma$ = 2 power law and nominal absorption.
It is roughly consistent
with a $\Gamma$ = 1 power law.

Source p5 is coincident with a small angular size galaxy in the
DSS image (see Figure 13), named ANON 3 in \citet{s00}.
No redshift is available for this object, so it is unclear at present
whether it is a background galaxy or a dwarf galaxy in the NGC 4410
group.  
Of the remaining sources, source p2 
appears to be associated with an optical concentration
in the southeastern stellar tail
(see Figure 13), p4 lies in the A/B$-$C bridge near the bright bridge source b1,
p6 is north of the star forming ring,
and p7 may be associated with the northwestern tail of NGC 4410A.
Source p1 is closest to the galactic nuclei, at a projected distance 
of 27 kpc from the nucleus of NGC 4410A.
Sources p1, p4, and p6, which are in the field of view of
the H$\alpha$ map of \citet{d02}, are not associated with
H~II regions.  The southeastern and northwestern tails, including
sources p2, p3, p7, and p8, were not in the field of view of
the \citet{d02} H$\alpha$ image.

In Figure 7, the 4.9 GHz radio contours of \citet{h86}
are superposed on the Chandra 0.3 $-$ 8 keV map.
Notice that the radio contours distinctly bend around source p1,
so we speculate that this object may be a radio source as well.
This possibility should be tested with higher resolution
radio data.
In addition, a radio point source at the location of p3
was found by \citet{h86} (named `Knot G' by \citet{h86}).
Although the contours in Figure 7 are not exactly coincident,
the coordinates obtained from high resolution observations
for Knot G by \citet{h86} do exactly coincide with p3.
At the time of the \citet{h86} radio observations,
Knot G had a flat/inverted radio spectrum with
$\alpha$ = $-$0.5 $\pm$ 0.3, consistent with some 
hard state galactic black hole candidates \citep{c00}.
As noted above, this source has a hard spectrum in
the Chandra data.

We did not detect an X-ray counterpart to 
the type I supernova SN 1965A,
which has a quoted position of 17$''$ east, 10$''$ north
of the NGC 4410B nucleus \citep{ks71, barbon91}.
Our upper limit of 
$\sim$10$^{39}$ erg~s$^{-1}$ for SN 1965A is consistent
with 
the upper limit to the X-ray luminosity
of the type Ia SN 1992A of 5 $\times$ 10$^{38}$
erg~s$^{-1}$ \citep{sp93}, and the measured
X-ray luminosity of the type Ic
supernova SN 2002ap, 
$\sim$10$^{38}$ erg~s$^{-1}$
\citep{sutaria03}.
Historical
type II supernova typically have X-ray luminosities
of $\sim$10$^{38}$ $-$ 5 $\times$ 10$^{39}$ erg~s$^{-1}$ 
(i.e., \citet{kaaret01}, \citet{schlegel96}).

\section{Discussion}

\subsection{The Galaxies}

The 0.3 $-$ 8 keV luminosities for NGC 4410B, C, and D are
5 $\times$ 10$^{39}$~erg~s$^{-1}$, 8 $\times$ 10$^{39}$~erg~s$^{-1}$,
and 10$^{40}$~erg~s$^{-1}$, respectively (Table 1).
From the POSS generation I plates \citet{cab99}
found blue O magnitudes of 16.3 and 16.5 
for NGC 4410C and D, respectively,
while \citet{deV93} gives a blue magnitude of B = 14.9 for 
NGC 4410B.  
Using the relation between O and B magnitudes shown in
\citet{cd02} and M$_{B}$($\sun$) = 5.48 gives L$_B$ = 1.6 $\times$
10$^{10}$ L$_{\sun}$, 9.8 $\times$ 10$^9$ L$_{\sun}$,
and 8.7 $\times$ 10$^9$ L$_{\sun}$ for NGC 4410B, C, and D.
These luminosities
are consistent with the \citet{f92} L$_X$/L$_B$ relationship for normal
spiral galaxies. 

The galaxy ANON 3,
which has a flattened appearance on the DSS images, has
an O magnitude of 19.3 \citep{cab99}.
Thus, if it is at the distance of NGC 4410, it has a blue
magnitude of only 4.9 $\times$ 10$^8$ L$_\sun$.  This
implies an X-ray
excess of a factor of 14 above the \citet{f92} relation
for normal spirals.
This is unexpectedly large for such a low blue luminosity galaxy,
suggesting that instead it may be a 
background galaxy, perhaps with an active nucleus.

\subsection{The Intragroup Emission and the Group Evolutionary State}

The luminosity of the intragroup medium, 10$^{41}$ erg~s$^{-1}$,
is at the lower end of the range for the
elliptical-dominated groups studied by \citet{mz98},
which typically have L$_x$ $\sim$ 10$^{41}$ $-$ 5 $\times$ 10$^{42}$~erg~s$^{-1}$.
This supports the idea that the NGC 4410 group is in the process of evolving from
a spiral-dominated group (which typically have 
no X-ray-emitting intragroup medium)
to an elliptical-dominated group.
The Chandra X-ray surface brightness
gives an average density of this intragroup gas of 2 $\times$ 10$^{-4}$
cm$^{-3}$, assuming
standard cooling functions \citep{mc77, mccray87}, a temperature of 0.75 keV,
and 
a spherical distribution with a radius of 78 kpc.
The total mass of this component is
$\sim$1.1 $\times$ 10$^{10}$ M$_{\sun}$.

The galaxies in the inner portion of the NGC 4410 group
appear to be somewhat deficient in HI compared to their blue
luminosities and morphological types (Smith 2000).  
This HI deficiency is likely due to gravitational interactions
between the galaxies, rather than ram pressure stripping by
an intracluster medium.  The morphology of the HI in the group
is consistent with tidal stripping (Smith 2000).
Assuming 
the galaxies are moving at the luminosity-weighted 
velocity dispersion in the group of 225 km s$^{-1}$ \citep{s00},
an estimate
of the intragroup medium density required to ram pressure
strip the HI based on the \citet{gg72} criteria
is about 100 times larger than the density inferred 
from the Chandra observations.

\citet{v01}
conducted 
a large HI survey of compact groups, and 
found significant
HI deficiency compared to the field.  As we found
for NGC 4410, for their groups \citet{v01} concluded the HI
deficiency was likely due to interactions rather than stripping
by ram pressure.
They suggest an evolutionary scenario for groups, in which
interactions create HI-rich tidal features, while depleting the
inner galaxies of HI.   
The HI in the tails is then ionized,
causing the group as a whole to become deficient in HI.
The HI in the NGC 4410 group is mostly found in tidal features
\citep{s00},
putting NGC 4410 in a later stage in this evolutionary
sequence.

The galaxies in the inner part of
the NGC 4410 group are likely in the process of merging to form
a central dominant elliptical.  The peculiar Sab galaxy NGC 4410A
is by far the most luminous galaxy in the group at optical
wavelengths, contributing $\sim$25$\%$ of the blue luminosity,
and having an absolute blue magnitude of $-$21.1, approaching
that of M87 \citep{s00}. 
At present, NGC 4410A contains 4 $\times$ 10$^9$~M$_{\sun}$
of molecular gas and $\sim$10$^9$~M$_{\sun}$ of atomic hydrogen \citep{s00}.
As the system evolves, it is likely that much of this 
neutral gas will be used
up in star formation or ionized by star formation, shocks,
or the active nucleus.
If this gas is converted into X-ray-emitting intragroup gas,
it will boost the X-ray luminosity to $\sim$1.5 
$\times$ 10$^{41}$~erg~s$^{-1}$.
After the final merger, this system may resemble the isolated
elliptical NGC 1132, which is surrounded by an X-ray halo
\citep{mz99}. 
Since clusters of galaxies appear to form hierarchically,
from the accretion of small groups \citep{bing93, zz95},
it is likely that the central dominant galaxies now seen
in galaxy
clusters originally formed in groups by a process similar to
that now occuring in NGC 4410.

In some of its properties the NGC 4410 group resembles the groups
HCG 16 
and HCG 92 (Stephan's Quintet).
All three of these groups contain strongly interacting galaxies,
with
tidal tails and bridges,
most of which are classified as spirals or S0s in NED.
Like the NGC 4410 group,
both HCG 16 and NCG 92 have X-ray-emitting intragroup
gas, with similar X-ray luminosities, temperatures,
and inferred abundances \citep{b03, t03}.
There are some differences between these groups, however.
HCG 16 has a higher HI mass than the inner part
of the NGC 4410 group, with
M$_{HI}$ $\sim$ 4 $\times$ 10$^{10}$~M$_{\sun}$, and HI is present
in both tidal features and in the inner parts of the galaxies
\citep{v01}.
This suggests that HCG 16 is in an earlier evolutionary stage
than the NGC 4410 group.
Stephan's Quintet has more HI than the NGC 4410 inner
group, with about 10$^{10}$~M$_{\sun}$ \citep{v01}, 
although, like NGC 4410 but
unlike HCG 16,
the HI in Stephan's Quintet is mostly in the tidal features.
Thus Stephan's Quintet may be in an evolutionary stage between
that of HCG 16 and the NGC 4410 group.

There are also differences in X-ray morphology.
In
Stephan's Quintet,
much of the diffuse X-ray-emitting gas
lies in a narrow north-south ridge running through the group
\citep{t03}.  \citet{t03} suggests that this feature is a shock front
caused by a high velocity intruder galaxy entering the group.
The diffuse gas in the NGC 4410A group is somewhat elongated,
but not to the extent seen in Stephan's Quintet, while the 
gas in HCG 16 has a more relaxed appearance \citep{b03}.

Of these three groups, only NGC 4410 contains a radio galaxy with
radio lobes that extend beyond the optical galaxy.
However, the optically-brightest galaxies in HCG 16 and
Stephan's Quintet are both Seyfert galaxies \citep{r96, h82}.
NGC 7319, the brightest galaxy in Stephan's Quintet, has a small (2 kpc diameter) 
triple radio structure similar to those of
FRII radio galaxies, with a flat-spectrum core and two asymmetric
extended lobes \citep{x02}.  This structure may
be due to the interaction of radio jets
with the interstellar medium in NGC 7319 \citep{x02}.
Thus NGC 7319 may eventually evolve into a classical double-lobed radio galaxy.

The NGC 4410A radio lobes may
have played a role in the evolution of the group as a whole.
They may have been disturbed by the interstellar medium
during the gravitational interaction, and 
may in turn have affected the gas around them.
As discussed in \citet{s00},
the density of intragroup
medium in the NGC 4410 group is likely too small to account
for the distortion of the radio lobes, implying that the distortion
is probably due to the interstellar material.
Radio jets may also contribute to heating the interstellar/intracluster
medium \citep{begel01, rhb02}.

The existence of large radio lobes
in the merging galaxy NGC 4410A shows that such structures can
occur fairly early in the formation of an elliptical, when it 
is not yet an elliptical.
NGC 4410A thus stands in contrast to the three `compact symmetric
object' radio galaxies 
studied
by \citet{p01}, which have normal outer optical isophotes
but significant ellipticity and position angle variations in the inner region.
In these three galaxies, the optical morphology suggests that a major merger
occurred $\ge$10$^8$ years ago, while 
the radio morphology indicates a recent (10$^3$ $-$ 10$^4$ years) 
turn-on of jet activity, implying a time delay between such a merger
and jet activity.
This may mean that minor mergers, which will not drastically
alter the outer isophotes, may also trigger jet activity,
or that 
jet activity is intermittent in 
elliptical galaxies, as suggested by 
\citet{rb97} and \citet{mcn01}, or that jets can also
be present in spirals but are generally smothered by dense
interstellar clouds \citep{wang00}, as suggested for NGC 7319 \citep{x02}.
Once formed, jets may 
affect the evolution of the interstellar and
intragroup gas,
providing periodic heating to
the interstellar and intracluster/intragroup gas.

\subsection{The NGC 4410A Ring and the Diffuse X-Ray Emission 
in NGC 4410A}

The peculiar ring-like structure visible in optical
images of NGC 4410A (Figure 2, \citet{d02}) is reminiscent
of those in classical ring galaxies
\citep{lt76, ts77}.
The existence of luminous H~II regions along the ring
is also consistent with a collisional ring scenario, as such
H~II regions are often found in classical ring galaxies
(e.g., \citet{h95}).
However, the 
optical spectrum
of the western edge of the NGC 4410 ring 
is indicative of shocks \citep{d02},
which 
is not typical of ring galaxies \citep{j86, b98}.

The detection of extended x-ray emission along the
western half of the NGC 4410 ring provides another
test of the collisional-ring hypothesis.
NGC 4410 is not the only ring(-like) galaxy with 
X-rays detected along the optical ring. 
ROSAT images of the Cartwheel 
\citep{wti99} show X-rays along the southwestern
portion of the ring, with an X-ray luminosity
of 2 $-$ 3 $\times$ 10$^{41}$ erg s$^{-1}$.
Higher resolution Chandra observations of the Cartwheel show
that this X-ray emission is largely
due to point sources, with about 10$^{40}$ erg~s$^{-1}$
being due to diffuse gas \citep{wt03}.
This is similar to 
our measurement of 
L$_X$ = 6 $\times$ 
10$^{39}$ erg s$^{-1}$ for the ridge
of X-ray emission along the western
part of the NGC 4410 ring.
However,
in contrast to NGC 4410,
in the Cartwheel the part of the ring brightest
in x-rays also has the most star formation
\citep{h95, wti99}, while in NGC 4410, the 
western part of the ring is detected in X-rays,
and the H~II regions are located in the
eastern part (Figure 2, \citet{d02}).

The extended X-ray emission in both the Cartwheel and NGC 4410
is unlikely to be due to 
shocks from an expanding density wave caused by a 
collision with another galaxy.
Typical expansion velocities for collisional rings are
50 $-$ 100 km s$^{-1}$ \citep{sh93, g96}.
As discussed by \citet{wti99},
such velocities imply very low gas temperatures 
of $\sim$50 eV.  Such low temperatures would cause strong
line emission in the extreme ultraviolet, rather than
a dominant X-ray continuum \citep{mccray87}.

Although shocks from an expanding density wave are not
likely to have caused the diffuse X-rays directly, they may
be indirectly responsible, arising as a consequence of
star formation triggered
by an expanding ring.
Alternatively, star formation triggered by another process
may be responsible for the diffuse X-rays in this system.
The total X-ray luminosity in the 
diffuse component in 
NGC 4410A+B, compared to other star formation indicators, is consistent with
that expected from star formation.
The far-infrared luminosity of NGC 4410A+B is
3.9 $\times$ 10$^9$ L$_{\sun}$ \citep{s00}.
Comparison with the total diffuse X-ray emission in NGC 4410A+B
of 6 $\times$ 10$^{40}$ erg~s$^{-1}$
gives a ratio consistent with the
\citet{djf92} and \citet{ss98}
L$_X$ vs. L$_{FIR}$
relationships for star forming galaxies, thus NGC 4410 does not have
a large excess of diffuse X-ray emitting gas compared to normal
star forming galaxies.
Another useful comparison is with the 
non-nuclear H$\alpha$ luminosity of NGC 4410A+B,
1.7 $\times$ 10$^{41}$ erg s$^{-1}$
\citep{d02}.  The diffuse L$_X$/L$_{H\alpha}$ $\sim$ 0.35 is
consistent with global values for starburst galaxies
\citep{poc96}.
The total diffuse X-ray luminosity in NGC 4410A is
similar to the soft X-ray luminosity found in the prototypical
spiral-to-elliptical merger NGC 7252 \citep{amt02}.

The extended x-ray emission in NGC 4410A (and in
other star-forming galaxies) is not likely
to be due to supernova remnants alone.
The typical X-ray luminosity of a 
supernova remnant
is $~$10$^{36}$ erg s$^{-1}$, while its lifetime is
about 2 $\times$ 10$^4$ years \citep{c81}.
Comparison with the observed X-ray luminosity of the NGC 4410 ring
implies a supernova rate of 0.3 per year, if all the
X-ray emission in the ridge
is due to supernovae.
If the total diffuse X-ray emission in NGC 4410A+B,
L$_X$ $\sim$ 6 $\times$ 10$^{40}$ erg s$^{-1}$,
is assumed to be due to supernovae, the implied supernova
rate is even higher, at 3 supernovae per year.
These implied rates are high compared to the total
rate of star formation in NGC 4410A,
1 $-$ 4 M$_{\sun}$~year$^{-1}$ \citep{d02} (which in any
case is mainly observed in the eastern portion of NGC 4410A,
not along the western arc).
Expected supernova rates, relative to star formation rates,
are $\sim$0.03 supernova/(M$_{\sun}$~year$^{-1}$) \citep{t01}, thus, based
on the non-nuclear H$\alpha$ luminosity, we expect only 0.03 $-$ 0.12 supernova
per year
in NGC 4410A.
In general, supernova remnants are believed to contribute
only a small amount to the X-ray emission from starburst
galaxies \citep{pr02}.

A related phenomenon more likely to be responsible
for the diffuse X-ray emission is the combined effect
of stellar winds from massive stars and supernovae,
creating an expanding bubble of hot gas.
Expansion of the bubble may shock the surrounding gas,
contributing to the X-ray radiation.
Such superbubbles in starburst
galaxies may be capable of generating up to L$_X$ $\sim$
10$^{39-41}$ erg~s$^{-1}$ \citep{ss98}.
Chandra observations of the starburst galaxy NGC 253
\citep{strick00} reveal a 
$\sim$0.5 kpc radius
conical-shaped outflow from the 
nucleus, with an
edge-brightened X-ray morphology similar to that seen
in H$\alpha$.
\citet{strick00} conclude that the diffuse X-rays from NGC 253
originate from an interface region between the wind and the ambient
interstellar medium.
In the SBc galaxy NGC 3079,
a 0.6 kpc radius superbubble is seen,
with matching X-ray and optical line morphologies
\citep{chv02}.  The X-ray luminosity associated
with this bubble is 3 $\times$ 10$^{42}$ erg~s$^{-1}$, 50 times higher
than the total extended X-ray luminosity of NGC 4410A+B.
In NGC 3079, unlike in NGC 253, the X-rays are not significantly
edge-brightened, and the cause of the superbubble (AGN or starburst) is not
clear.

The western arc/X-ray ridge in NGC 4410, however, is much
larger than the superbubbles in these
galaxies, extending $\sim$4 kpc from the nucleus.
The expansion timescale of a superbubble can be estimated from
$ t_7^{-3/5} \sim 2.8 L_{mech}^{1/5} n_H^{-1/5} / r_{shell} $,
where $t_7$ is the timescale in units of 10$^7$ years, L$_{mech}$ is the
mechanical energy of the supernova in units of 10$^{43}$ erg~s$^{-1}$,
n$_H$ 
is the average hydrogen number density in units of 1~cm$^{-3}$, and r$_{shell}$
is the radius of the bubble in kpc
\citep{shull95,t01}.
Using
L$_{mech}$ $\sim$ $\eta$E$_{SN}$R$_{SN}$, where $\eta$ $\sim$ 0.1 is the efficiency
of the kinetic energy deposited into the gas \citep{dw80}, 
E$_{SN}$ $\sim$ 10$^{51}$ erg
is the energy of a single supernova, R$_{SN}$ is the number of supernova per year,
and the hydrogen number density n$_H$ $\sim$ 1 cm$^{-3}$
gives L$_{mech}$ $\sim$ 0.9~$-$~3.9~$\times$~10$^{41}$~erg~s$^{-1}$,
using the supernova rate implied by the H$\alpha$ luminosity.
This implies a timescale of 5 $-$ 9 $\times$ 10$^{7}$ years
for the bubble.

If the X-ray/optical emission-line arc in NGC 4410 was indeed caused by
a superbubble, it might be caused by the active nucleus rather than
by star formation.
Theoretical models of a radio jet impacting the interstellar
medium or intragroup medium predict an X-ray shell
surrounding the radio lobes \citep{rhb01}.  Such shells
are observed
in a number of radio galaxies \citep{cph94, b93, hs98}.
Unfortunately, at present no high spatial resolution, high 
sensitivity radio observations are available for NGC 4410A
for comparison with the Chandra data.
In the 1986 maps of \citet{h86}, only the southern radio
lobe is visible.  
In the higher sensitivity but low spatial resolution
map of \citet{s00}, a second lobe is visible,
but its detailed morphology is uncertain.
Thus at present it is unclear whether or not the X-ray ridge
in NGC 4410A bounds a radio lobe.

\subsection{The Point Sources}

As discussed above,
of the eight point sources found in the NGC 4410 field,
one of them (p5) is associated
with the faint optical galaxy ANON 3, and 
may be 
a background source.  
The second source with an optical
counterpart, p2, may also be a background source.
The optical counterpart is too faint to have been included in
the \citet{cab99} catalog of POSS I sources, however,
it is clearly visible on the POSS II plates.
Roughly calibrating these images by comparison with
the \citet{cab99} galaxy magnitudes,
we estimate B $\sim$ 19.5 and V$\sim$ 19 for the optical
counterpart to p2.
At the distance of NGC 4410, this corresponds to 
a blue luminosity of 2.4 $\times$ 10$^8$ L$_{\sun}$ 
(M$_B$ $\sim$ $-$15.4 and M$_V$ $\sim$ $-$15.9).
This is much more luminous in the visible than almost all of the 
so-called `super star clusters' found in star-forming
galaxies ({\it e.g.}, \citet{holtz92, holtz96}).
This source has a log(F$_X$/F$_V$) ratio of $-$1.4 (calculated
as in \citet{m88}), much smaller 
than that of the apparent optical
counterparts to some Ultraluminous X-ray sources (ULXs) 
found in other galaxies.
For example, the apparent optical counterparts to ULXs found
in M82 and NGC 4565
have M$_B$ $\sim$ $-$5 and log(F$_X$/F$_V$) $\sim$ 
2
\citep{swk91, wu02}.

Thus p2 is clearly not in the same class as these ULXs,
being much brighter in the optical (if the POSS
optical source is indeed related to the X-ray source).
It is more likely to be a background active galaxy.
Although luminous in the optical compared to these other
ULXs, it is faint in the optical relative to the X-ray compared to normal galaxies;
it lies above the \citet{f92} L$_X$/L$_B$ relation for normal
galaxies.
If it is at the distance of NGC 4410, its optical 
luminosity is typical of a dwarf galaxy.  As argued earlier for ANON 3,
it is unlikely 
that such a low luminosity galaxy
would harbor an active galaxy.   
Thus p2 may also be a background object.

The remaining six candidate ULXs without optical counterparts on
the POSS images have lower limits to log(F$_X$/F$_V$) 
between $-$1.5 to $-$0.4,
assuming an optical limit of V $\sim$ 20.5 from the POSS I plates.
These limits are not sensitive enough to distinguish between
ULXs and background
active galaxies, via 
comparison with typical values for BL Lacs and other AGNs ({\it i.e., } \citet{stocke91}).

As discussed in Section 3.3, 
two of our candidate ULXs, p3 and p8, have 
hard X-ray spectra compared
to a $\Gamma$ = 2 power law.
Candidate ULXs in other galaxies
typically have spectra consistent with $\Gamma$ $\sim$ 2 
power laws \citep{r01, lira02}, though there is some evidence
for harder spectra ($\Gamma$ $\sim$ 1.2) among 
the more luminous ULXs \citep{z02}.
Thus p3 and p8 may also be
background sources (Seyfert 2, for example).
Alternatively, they may be `microquasars', particularly
p3, whose large radio flux
may indicate that we are viewing nearly straight down the jet
axis of a `microquasar' such 
as GRS~1915+105.
Intriguingly, p3 and p8 are spatially located
together
southeast of NGC 4410A
near p2, which has an optical counterpart.
If these are indeed all background objects, they
may be part of a background cluster.
Alternatively, they may be ULXs associated with
the NGC 4410 tidal tail.

Thus, our four best candidate ULX sources are
p1, p4, p6, and p7.
Given the star formation rate estimate for NGC 4410A of 
1~$-$~4~M$_{\sun}$~year$^{-1}$ \citep{d02},
we expect to find 1 $-$ 4 ULXs associated with NGC 4410A,
if ULXs follow a universal luminosity function dependent
upon star formation rate, as suggested by \citet{g03}.
Our detection of four possible candidates in NGC 4410 is consistent
with this expectation.
As noted above, however, three of
these sources (p1, p4, and p6)
are not associated with known star formation regions.

The possible association of radio emission with one 
of these candidate ULXs, p1, is intriguing.  To date, only
one other ULX with a possible radio counterpart has been found \citep{k03}.
If these objects are
indeed radio-bright ULXs, this suggests that they are not
intermediate mass black holes, but instead beamed stellar mass objects
\citep{k03}.
Alternatively, p1 may be a background radio-loud active galaxy.

\section{Summary}

The peculiar galaxy NGC 4410A contains a X-ray-bright
active nucleus as well as diffuse extended emission, including
a ridge of X-ray emission coincident with optical line emission.
This ridge has an optical spectrum indicative
of shocks, suggesting
it may be the rim of an expanding superbubble caused by
either the on-going star formation in NGC 4410A or its active
nucleus.  Outside of the optical galaxies in the group,
a hot intragroup gas with L$_X$ $\sim$ 10$^{41}$ erg~s$^{-1}$ is observed,
at the lower edge of the range observed for 
elliptical-dominated groups.  This suggests that the NGC 4410
group is in the process of evolving from a spiral-dominated group
into an elliptical-dominated group.
Four or five possible ULXs are seen in the Chandra maps of NGC 4410A,
one of which has a possible radio counterpart.  Such a radio
counterpart argues against the intermediate mass black hole
model for this source, and supports the hypothesis that this
is simply a beamed stellar-mass binary.

\acknowledgements

We thank the Chandra team for making this research possible.
This research has made use of the NASA/IPAC Extragalactic Database
(NED) which is operated by the Jet Propulsion Laboratory, California
Institute of Technology, under contract with the National Aeronautics
and Space Administration.
We appreciate the helpful suggestions of the anonymous referee.
We thank Anna Wolter for helpful comments on the manuscript,
and for providing information on the Cartwheel
Chandra observations
in advance of publication.
We also thank Kathy Manning and John Houck for advice with the data reduction
software.
We are pleased to acknowledge funding for this project from a
NASA General Observer Chandra grant.
M.A.N. was supported by NASA grant NAS8-01129.

\clearpage


\figcaption{\small An optical R-band image of the inner part of the NGC 4410
galaxy group, obtained with the South Eastern Association for Research in
Astronomy (SARA) 0.9\,m telescope (from Smith 2000, Donahue et al. 2002).  
The stretch is set to emphasize the
faint bridges connecting NGC 4410 A+B with NGC4410 C and 
NGC 4410 D, as well as the
long tail extending to the southeast of NGC 4410 A+B.  The field of view of
this image is $\approx 6.7$\,arcmin $\times 6.3$\,arcmin (slightly
smaller than the ACIS-S S3 chip).  North is up and east is to the
left.}

\clearpage 

\figcaption{\small 
A closer view of NGC 4410 A+B (from
Donahue et al. 2002).  The left image is a red
continuum image, while the right image is an H$\alpha$+[N~II] image.  The
field of view is $\approx0.8$\,arcmin $\times 1.1$\,arcmin. NGC 4410A, the
galaxy on the right in these images, has a prominent bulge surrounded by a
ring-like or loop-like structure.  Optical spectroscopy (Donahue et 
al. 2002) confirms that the bright clumps in the ring and the extremely
bright knot southwest of the NGC 4410A nucleus are extremely luminous H~II
regions, however, the arc to the northwest of the NGC 4410 nucleus
has an optical spectrum indicative of shocks.  
The most luminous H~II region is twice as luminous as 30
Doradus.
The registration on these images has been adjusted to match the 
Chandra and 
Hummel et al. (1986) radio continuum positions of the
nucleus of NGC 4410A.
}

\figcaption{\small 
The final Chandra 0.3 $-$ 8 keV image (greyscale), superposed on
the optical Digitized Sky Survey (DSS) image (contours).
Note the X-ray counterparts to NGC 4410A+B, NGC 4410C, and NGC 4410D,
as well as numerous background sources.
This map was made using an exposure map that was weighted 
using the spectrum of a 57\farcs5 $\times$ 40$''$
region enclosing the NGC 4410A+B complex.
Note that the morphology of the extended emission does not vary
with different assumptions used in making the exposure map.
}

\figcaption{\small 
The low energy (0.3 $-$ 1 keV) Chandra image of the
inner portion of the NGC 4410 group (greyscale).
This map was made using a monochromatic 0.8 keV exposure map.
As in the previous
figure, the contours are 
from the optical Digitized Sky Survey (DSS) image.
Note the possible faint X-ray counterpart to NGC 4410F.
}

\figcaption{\small 
The middle energy (1 $-$ 2.5 keV) Chandra image of the
inner portion of the NGC 4410 group (greyscale).
This map was made using a monochromatic 1.5 keV exposure map.
As in the previous
figures, the contours are 
from the optical Digitized Sky Survey (DSS) image.
}

\figcaption{\small 
The high energy (2.5 $-$ 8 keV) Chandra image of the
inner portion of the NGC 4410 group (greyscale).
This map was made using a monochromatic 3 keV exposure map.
As in the previous
figures, the contours are 
from the optical Digitized Sky Survey (DSS) image.
}

\figcaption{\small 
The Hummel et al. (1986) 4.9 GHz radio continuum map, superposed
on the 0.3 $-$ 8 keV Chandra map.
}

\figcaption{\small 
The Chandra 0.3 $-$ 8 keV image of the NGC 4410A+B region.
The bright source is centered on the nucleus of NGC 4410A,
while the secondary source to the east is associated with the nucleus
of NGC 4410B.  The extension to the northwest is coincident with
the shock-ionized optical emission line gas seen in the H$\alpha$
map (Figure 2).
}

\figcaption{\small 
Left: A broadband red archival Hubble Space Telescope image.
Right:
The HST red image (contours), superposed on
the unsmoothed Chandra 0.3 $-$ 8 keV image of the NGC 4410A+B region.
Note that the optically-bright knots to the southeast and northeast
of the nucleus of NGC 4410A are not visible as discrete sources
in the X-ray map.  The extended X-ray radiation to the northwest
is concentrated along the inner edge of the optical arc.
}

\figcaption{\small 
The Chandra 0.3 $-$ 8 keV image of the NGC 4410A+B region (greyscale),
with the Donahue et al. (2002) 
H$\alpha$+$[$N~II$]$ image superimposed
(contours).
The 
H$\alpha$+$[$N~II$]$ image has been re-registered so that the NGC 4410A
nucleus is coincident with the radio continuum and X-ray nucleus.
Note the $\sim$3$''$ extension to the southwest of the NGC 4410A nucleus
which is present in both the X-ray and in the optical line emission.
Note also that the concentration of X-ray emission 18$''$$-$20$''$ 
to the northwest of the NGC 4410A nucleus is coincident with the
arc of ionized gas seen in the optical image.
In addition, note that the HII regions to the southeast and northeast
of the nucleus of NGC 4410A are not visible as discrete sources
in the X-ray map.  
}

\figcaption{\small 
a) The 0.5 $-$ 5 keV spectrum of the inner
1$''$ of NGC 4410A (solid line).
The dotted line is the best fit to an absorbed power law,
with the column density fixed to the nominal value.
A pile-up model was assumed.
b)
The 0.5 $-$ 5 keV spectrum of the 1$''$ $-$ 
5$''$ annulus around NGC 4410A (solid line).
The dotted line is the fit to a power law plus Mekal function,
as described in the text (see Table 1).
c)
The 0.5 $-$ 5 keV spectrum of the 5$''$ $-$ 
10$''$ annulus around NGC 4410A (solid line).
The dotted line is the fit to a power law plus Mekal 
function (see text and Table 1).
d)
The 0.5 $-$ 5 keV spectrum of the bright bridge source.
The dotted line is the fit to a power law 
function, as described in the text (see also Table 1).
}

\figcaption{\small 
a) The observed 0.5 $-$ 5 keV spectrum of the diffuse intragroup gas in
the NGC 4410 group (solid line), compared to expected background counts
from the deep Markevitch observations (dotted lines).
b) The background-subtracted spectrum for the diffuse gas in the NGC 4410
group (solid line), compared to the best-fit model (dotted line).
}

\figcaption{\small 
An overlay of the positions of the candidate Ultraluminous
X-Ray sources (ULXs) (see Table 2), on the red Second Generation
Digitized Sky Survey Image.  
}

\clearpage

\begin{deluxetable}{crrrrrrrrrrr}
\tabletypesize{\scriptsize}
\tablecaption{Spectral Fits to X-ray Sources in the NGC 4410 Field. \label{tbl-1}}
\tablewidth{0pt}
\tablehead{
\colhead{Source} &
\colhead{Function} &
\colhead{$\chi^2$/DoF} &
\colhead{${\rm n}_{\rm H}$}  & 
\colhead{$\Gamma$} & 
\colhead{kT} &
\colhead{L$_X$} \\
\colhead{}   &
\colhead{} &
\colhead{} &
\colhead{$\times$ 10$^{20}$ cm$^{-2}$}  & 
\colhead{} &
\colhead{} &
\colhead{(0.3$-$8 keV)} \\
\colhead{}   &
\colhead{} &
\colhead{} &
\colhead{}  & 
\colhead{} &
\colhead{} &
\colhead{(10$^{40}$ erg/s)} \\
}
\startdata
A ($<$1$''$)$^a$&Power Law&15/30&5$^b$&1.96$\pm$0.10&&25\\
A ($<$1$''$)$^a$&Power Law&14/30&3.7$\pm$$^{1.9}_{1.7}$&1.96$^b$&&23\\
A ($<$1$''$)$^a$&Power Law&18/30&6.7$\pm$$^{1.9}_{1.8}$&2.1$^b$&&22\\
A ($<$1$''$)$^a$&Power Law&27/30&10.7$\pm$$^{1.9}_{1.8}$&2.3$^b$&&20\\
A (1$''$$-$5$''$)&Mekal&2.2/9&5$^b$&&0.62$\pm$$^{0.08}_{0.07}$&4.1\\
A (5$''$$-$10$''$)&Mekal&1.3/4&5$^b$&&0.54$\pm$$^{0.12}_{0.14}$&1.9\\
B&Bremsstrahlung&$-$&5$^b$&&0.2$^b$&0.7\\
C&Bremsstrahlung&$-$&5$^b$&&0.5$^b$&0.4\\
D&Bremsstrahlung&$-$&5$^b$&&0.5$^b$&0.8\\
b1&Power Law&1.2/6&5$^b$&2.2$\pm$$^{0.3}_{0.2}$&&3.6\\
Intragroup Gas&Mekal&23/19&1.7$^b$&&0.69 $\pm$ $^{0.15}_{0.16}$&10\\
 \enddata

\tablenotetext{a}{
Used pileup model with $\alpha$ = 0.5 and psffract = 0.95 fixed.}
\tablenotetext{b}{Parameter fixed.}

\tablecomments{Fits are from 0.5 $-$ 5 keV, except in the case
of the inner annulus around NGC 4410A and the diffuse intragroup 
gas, where 0.6 $-$ 1.8 keV was used, and the outer NGC 4410A annulus,
where the energy range was restricted to 0.6 $-$ 1.2 keV.
Before fitting, data were rebinned to 20 counts/bin.
Luminosities were calculated assuming a distance of 97 Mpc.
The best-fit abundances for the inner and outer annuli of NGC 4410A
and the diffuse gas are
0.06 $\pm$ $^{0.06}_{0.03}$, 0.11 $\pm$ $^{1.16}_{0.09}$,
and 0.025 $\pm$ $^{0.025}_{0.024}$ solar,
respectively.
}

\end{deluxetable}

\clearpage

%
%
\begin{deluxetable}{lrrrrrrrrrcrrrccccr}
\tabletypesize{\scriptsize}
\def\et#1#2#3{${#1}^{+#2}_{-#3}$}
\tablewidth{0pt}
\tablecaption{Counts and Luminosities of Point Sources \label{tab:points}}
\tablehead{
\colhead{Source} &
\multicolumn{3}{c}{R.A.}&
\multicolumn{3}{c}{Dec.}
& \colhead{$E_1$} & \colhead{$E_2$} &
\colhead{$E_3$\tablenotemark{a}} &  \colhead{$E_A$} & \colhead{$E_B$} & 
\colhead{$E_C$\tablenotemark{b}} & $\quad$ &\colhead{$L_A$} & \colhead{$L_B$} & 
\colhead{$L_C$} & \colhead{Total\tablenotemark{c}} \\
&
\multicolumn{3}{c}{(J2000)}&
\multicolumn{3}{c}{(J2000)}
& \multicolumn{6}{c}{(Counts)} &&\multicolumn{4}{c}{($\times 10^{39}~ \rm erg~s^{-1}$)}}
\startdata
P1&12&26&31.857&+9&1&1.54 & 7 & 7 & 17 & 11 & 14 & 6 & &
   \et{1.4}{0.8}{0.6} & \et{3.6}{1.8}{1.4} & \et{6.1}{5.1}{3.3} &
   \et{11.1}{5.5}{3.6} \\
P2$^d$&12&26&34.266&+8&59&35.57& 6 & 1 & 8  & 7  & 6  & 2 & &
   \et{0.9}{0.7}{0.5} & \et{1.6}{1.3}{0.8} & \et{2.0}{3.5}{1.6} &
   \et{4.5}{3.8}{1.8} \\
P3$^e$& 12&26&32.457&+8&59&32.17 & 1 & 1 & 7  & 1  & 6  & 2 & &
   \et{0.1}{0.4}{0.1} & \et{1.6}{1.3}{0.8} & \et{2.0}{3.5}{1.6} &
   \et{3.7}{3.8}{1.8} \\
P4&12&26&32.429&+9&1&41.44 & 1 & 1 & 3  & 2  & 3  & 0 & &
   \et{0.3}{0.4}{0.2} & \et{0.8}{1.0}{0.5} & \et{0.0}{2.3}{0.0} &
   \et{1.0}{2.6}{0.6} \\
P5$^f$&12&26&32.792&+9&3&0.94 & 2 & 2 & 3  & 3  & 4  & 0 & &
   \et{0.4}{0.5}{0.3} & \et{1.0}{1.1}{0.6} & \et{0.0}{2.3}{0.0} &
   \et{1.4}{2.6}{0.7} \\
P6&12&26&29.354&+9&1&54.36 & 2 & 3 & 4  & 4  & 4  & 1 & &
   \et{0.5}{0.6}{0.3} & \et{1.0}{1.1}{0.6} & \et{1.0}{2.9}{0.9} &
   \et{2.6}{3.2}{1.2} \\
P7&12&26&22.985&+9&2&4.49 & 1 & 0 & 4  & 1  & 3  & 1 & &
   \et{0.1}{0.4}{0.1} & \et{0.8}{1.0}{0.5} & \et{1.0}{2.9}{0.9} &
   \et{1.9}{3.1}{1.1} \\
P8&12&26&33.458&+8&59&39.79 & 0 & 0 & 4  & 0  & 0  & 4 & &
   \et{0.0}{0.3}{0.0} & \et{0.0}{0.6}{0.0} & \et{4.1}{4.4}{2.5} &
   \et{4.1}{4.5}{2.5} \\ 
\enddata
\tablenotetext{a}{$E_1 =$ 0.5--0.95\,keV, $E_2 =$ 0.95--1.5\,keV, $E_3
=$ 1.5--8\,keV. A $\Gamma = 2$ power law source absorbed by a column
of $N_{\rm H} = 5 \times 10^{20}~{\rm cm^{-2}}$ should have equal
counts per energy bin.}
\tablenotetext{b}{$E_A =$ 0.5--1.4\,keV, $E_B =$ 1.4--3.4\,keV, $E_C
=$ 3.4--8\,keV. A $\Gamma = 2$ power law source absorbed by a column
of $N_{\rm H} = 5 \times 10^{20}~{\rm cm^{-2}}$ should have equal
flux per energy bin.}
\tablenotetext{c}{Absorbed luminosities, determined for each bin
individually, assuming a $\Gamma = 2$ power law
source absorbed by a column of $N_{\rm H} = 5 \times 10^{20}~{\rm
cm^{-2}}$ at a distance of 97 Mpc.}
\tablenotetext{d}{Possible optical counterpart on DSS.}
\tablenotetext{e}{Coincident with radio Knot G in Hummel et al. (1986)
radio continuum map.}
\tablenotetext{f}{Coincident with small angular size galaxy ANON 3 
on DSS image (as named in Smith 2000).}
\end{deluxetable}
%

\end{document}